\documentclass{ws-procs9x6}
\usepackage[british]{babel}
\usepackage{epsfig}

\begin{document}

\title{Numerical simulation with light Wilson-quarks}

\author{I.~Montvay}

\address{Deutsches Elektronen-Synchrotron DESY    \\
         Notkestr.\,85, D-22603 Hamburg, Germany  \\
         E-mail: istvan.montvay@desy.de}

\maketitle

\abstracts{
 The computational cost of numerical simulations of QCD with light
 dynamical Wilson-quarks is estimated by determining the
 autocorrelation of various quantities.
 In test runs the expected qualitative behaviour of the pion mass and
 coupling at small quark masses is observed.}

\section{Introduction}\label{sec1}
 In Nature there exist three light quarks ($u$, $d$ and $s$) which
 determine hadron physics at low energies.
 Numerical simulations on the lattice have to deal with them -- which
 is not easy because the known simulation algorithms slow down
 substantially if light fermions are involved.

 At present most dynamical (``unquenched'') simulations are performed
 with relatively heavy quarks, in case of Wilson-type lattice fermions
 typically at masses above half of the strange quark mass, and then
 chiral perturbation theory (ChPT)\cite{CHPT} is used for extrapolating
 the results to the small $u$- and $d$-quark masses.
 This extrapolation is better under controle if the dynamical quarks
 are as light as possible.

 In this talk I report on some recent work of the qq+q Collaboration
 concerning numerical simulations with light
 Wilson-quarks\cite{NF2TEST,BOSTON,PRICE}.
 We used the two-step multi-boson (TSMB) algorithm\cite{TSMB} which
 turned out to be relatively efficient for light fermions in previous
 investigations of supersymmetric Yang-Mills theory.
 (For a review with references see ref.\cite{SYMREV}).

\section{Estimates of computational costs}\label{sec2}
 In numerical Monte Carlo simulations the goal is to produce a sequence
 of statistically independent configurations which can be used for
 obtaining estimates of expectation values of different quantities.
 A measure of independence is provided by the values of the
 {\it integrated autocorrelation lengths} in the configuration sequence,
 usually denoted by $\tau_{int}^Q$.
 This depends on the quantity $Q$ of interest and gives the distance
 of statistically independent configurations.
\begin{figure}[ht]
\vspace*{-8mm}
\begin{center}
\epsfig{file=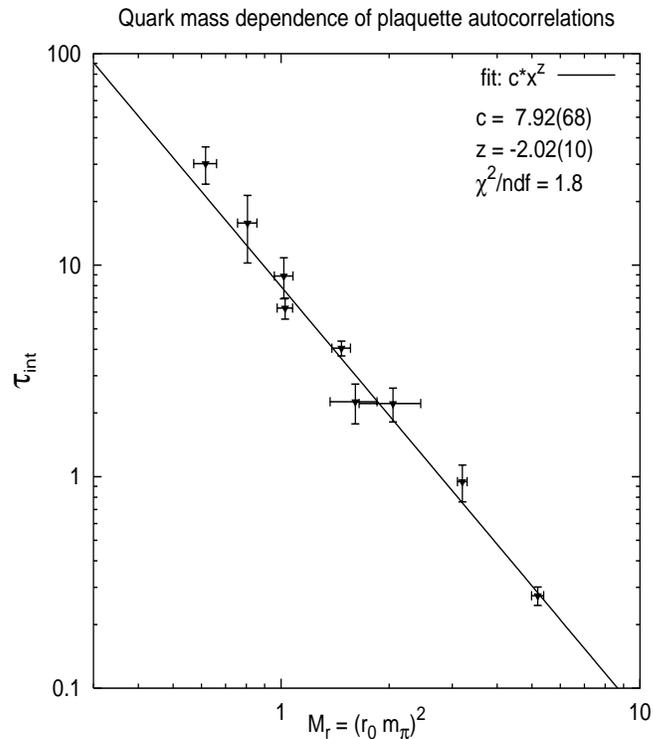,
        width=100mm,height=140mm,angle=-90,
        bbllx=50pt,bblly=-20pt,bburx=554pt,bbury=800pt}
\vspace*{-2mm}
\caption{\label{fig01}
 Power fit of the average plaquette autocorrelation given in units of
 $10^6 \cdot {\rm MVM}$ as a function of the dimensionless quark mass
 parameter $M_r$.
 The best fit of the form $c M_r^z$ is at $c=7.92(68),\; z=-2.02(10)$.}
\end{center}
\vspace*{-5mm}
\end{figure}

 The qq+q Collaboration has recently performed a series of test runs on
 $8^3 \cdot 16$, $12^3 \cdot 24$ and $16^4$ lattices with $N_f=2$ and
 $N_f=2+1$ quark flavours.
 The quark masses were in the range $\frac{1}{6}m_s < m_q < 2m_s$ and
 the autocorrelations of several quantities as average plaquette,
 smallest eigenvalue of the fermion matrix, pion mass and coupling etc.
 have been determined.
 The error analysis and integrated autocorrelations in the runs have
 been obtained using the {\em linearization method} of the ALPHA
 collaboration\cite{ALPHA:BENCHMARK}.

 The computational cost of obtaining a new, independent gauge
 configuration in an updating sequence with dynamical quarks can be
 parametrized, for instance, as\cite{NF2TEST}
\begin{equation}\label{eq01}
C = F\; (r_0 m_\pi)^{-z_\pi} \left(\frac{L}{a}\right)^{z_L}
\left(\frac{r_0}{a}\right)^{z_a} \ .
\end{equation}
 Here $r_0$ is the Sommer scale parameter, $m_\pi$ the pion mass, $L$
 the lattice extension and $a$ the lattice spacing.
 The powers $z_{\pi,L,a}$ and the overall constant $F$ are empirically
 determined.
 The unit of ``cost'' can be, for instance, the number of necessary
 fermion-matrix-vector-multiplications (MVMs) or the number of floating
 point operations to be performed.
 For an example on the quark mass dependence of the cost see figure
 \ref{fig01} which is taken from ref.\cite{NF2TEST}.
 This shows that the quark mass dependence in case of the average
 plaquette is characterized by a power $z_\pi \simeq 4$.
 For other quantities, as the smallest eigenvalue of the fermion matrix
 and the pion mass, a smaller power $z_\pi \simeq 3$ is observed.

 Other tests on the lattice volume and lattice spacing dependence showed
 a surprisingly mild increase in both directions if compared to the
 data\cite{NF2TEST} on $8^3 \cdot 16$ lattice at
 $a \simeq 0.27\, {\rm fm}$.
 Some results on the volume dependence of the average plaquette
 autocorrelation $\tau_{int}^{plaq}$ are given in the first four lines
 of table \ref{tab01}.
\begin{table}[ph]
\tbl{Runs for comparing the simulations costs (given in numbers of
 floating point operations) at different volumes and lattice spacings.}
{\footnotesize
\begin{tabular}{|c|c|c|c|r|}
\hline
label & lattice & $\beta$ & $\kappa$ & 
\multicolumn{1}{|c|}{$\tau_{int}^{plaq}\,[{\rm flop}]$}
\\ \hline
(e) & $8^3 \cdot 16$  & 4.76 & 0.190 & 4.59(37) $\cdot 10^{13}$
\\ \hline
(e16) & $16^4$        & 4.76 & 0.190 & 7.5(1.3) $\cdot 10^{14}$
\\ \hline
(h) & $8^3 \cdot 16$  & 4.68 & 0.195 & 1.7(6) $\cdot 10^{14}$
\\ \hline
(h16) & $16^4$        & 4.68 & 0.195 & 1.10(17) $\cdot 10^{15}$
\\ \hline
(E16) & $16^4$        & 5.10 & 0.177 & 2.1(4) $\cdot 10^{14}$
\\ \hline
\end{tabular}\label{tab01} }
\end{table}
 The runs with label (e16) and (E16) belong to almost the same quark
 mass ($M_r \simeq 1.4$) but have by a factor of about 1.5 different
 lattice spacing.
 A typical expectation for the power governing the lattice spacing
 dependence is $z_a=2$ which would imply by a factor of 2.25
 lager value for (E16) than for (e16).
 Compared to run (e) $z_L=4$ and $z_a=2$ would imply for (E16) an
 increase by a factor 36 instead of the actual factor $\simeq 4$.
 The observed relative gain is partly due to some improvements of the
 simulation algorithm (see ref.\cite{BOSTON}) and is, of course, very
 welcome in future simulations.
\begin{figure}[ht]
\vspace*{0mm}
\begin{center}
\epsfig{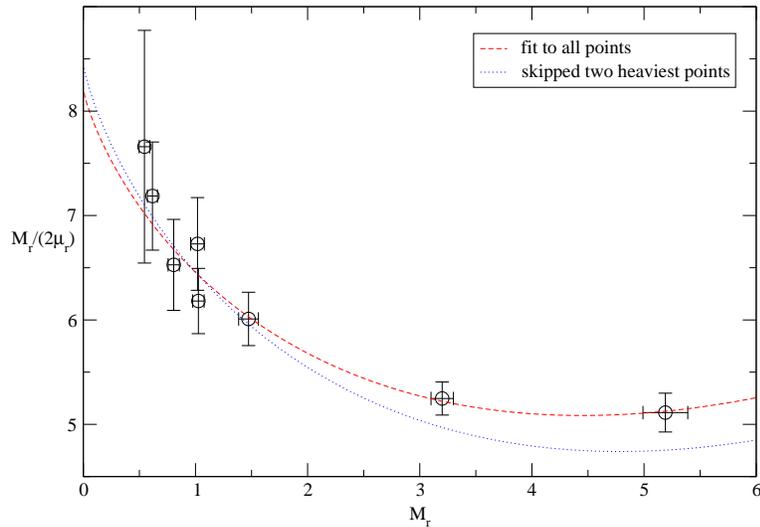}
\vspace*{-2mm}
\caption{\label{fig02}
 Fits of the pseudoscalar meson mass-squared with the one-loop ChPT
 formula.}
\end{center}
\vspace*{-5mm}
\end{figure}
\begin{figure}[ht]
\vspace*{-2mm}
\begin{center}
\epsfig{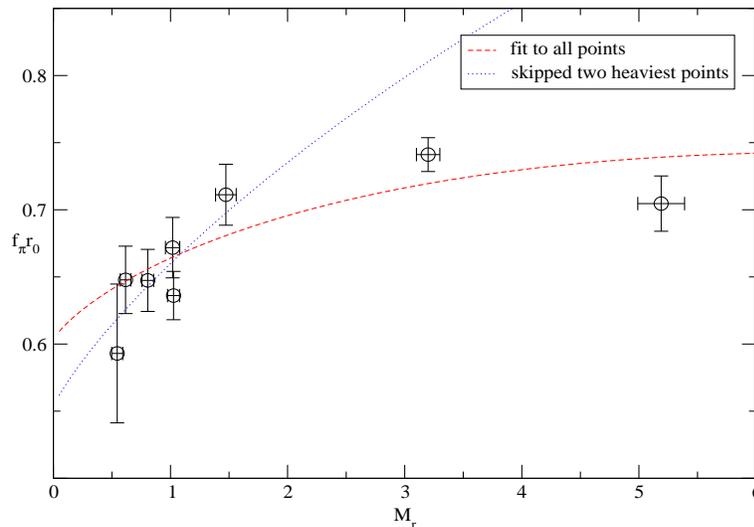}
\vspace*{-2mm}
\caption{\label{fig03}
 Fits of the pseudoscalar meson decay constant with the one-loop ChPT
 formula.}
\end{center}
\vspace*{-5mm}
\end{figure}

\section{Chiral logarithms?}\label{sec3}
 The behaviour of physical quantities, as for instance the pseudoscalar
 meson (``pion'') mass $m_\pi$ or pseudoscalar decay constant $f_\pi$
 as a function of the quark mass are characterized by the appearance of
 {\em chiral logarithms}.
 These chiral logs, which are due to virtual pseudoscalar meson loops,
 have a non-analytic behaviour near zero quark mass of a generic form
 $m_q\log m_q$.
 They imply relatively fast changes of certain quantities near zero
 quark mass which are not seen in present
 data\cite{CPPACS,UKQCD,JLQCD:CHLOGM,JLQCD:CHLOGF,CPPACS:Namekawa}.

 Although we have rather coarse lattices ($a \simeq 0.27\, {\rm fm}$)
 and, in addition, up to now we are working with unrenormalized
 quantities -- without the $Z$-factors of multiplicative renormalization
 -- it is interesting to see that the effects of chiral logs are
 qualitatively displayed by our data.
 Fits with the ChPT-formulas (see, for instance,
 ref.\cite{LEUTWYLER,DURR}) are shown in figures \ref{fig02} and
 \ref{fig03}.
 These are taken from ref.\cite{PRICE,GEBERT} where the fit parameters
 are also quoted.
 To see the expected qualitative behaviour with chiral logarithms in
 numerical simulations at small quark masses is quite satisfactory but
 for a quantitative determination of the ChPT parameters one has to go
 to smaller lattice spacings.



\end{document}